\documentclass[conference]{IEEEtran}
\IEEEoverridecommandlockouts
\usepackage{cite}
\usepackage{amsmath,amssymb,amsfonts}
\usepackage{algorithmic}
\usepackage{graphicx}
\usepackage{textcomp}
\usepackage{xcolor}

\usepackage{algorithm}
\usepackage{array}
\usepackage[caption=false,font=normalsize,labelfont=sf,textfont=sf]{subfig}
\usepackage{stfloats}
\usepackage{url}
\usepackage{verbatim}

\usepackage{tabularx}
\usepackage{multirow}
\usepackage{makecell}
\usepackage{booktabs}

\def\BibTeX{{\rm B\kern-.05em{\sc i\kern-.025em b}\kern-.08em
    T\kern-.1667em\lower.7ex\hbox{E}\kern-.125emX}}
\begin{document}

\title{CLF-ULP: Cross-Layer Fusion-Based Link Prediction in Dynamic Multiplex UAV Networks
}
\author{Cunlai Pu, Fangrui Wu, Zhe Wang and Xiangbo Shu,~\IEEEmembership{Senior Member,~IEEE}
\thanks{Manuscript received April 19, 2021; revised August 16, 2021.This work
 was supported in part by the National Natural Science Foundation of China
 under Grant 62427808. (Corresponding authors: Cunlai Pu, Xiangbo Shu.)}
\thanks{Cunlai Pu, Fangrui Wu, Zhe Wang and Xiangbo Shu are with
 the School of Computer Science and Engineering, Nanjing University of
 Science and Technology, Nanjing 210094, China (e-mail: pucunlai, wufangrui, zwang, shuxb@njust.edu.cn)}
}


\maketitle

\begin{abstract}
In complex Unmanned Aerial Vehicle (UAV) networks, UAVs can establish dynamic and heterogeneous links with one another for various purposes, such as communication coverage, collective sensing, and task collaboration. These interactions give rise to dynamic multiplex UAV networks, where each layer represents a distinct type of interaction among UAVs. Understanding how such links form and evolve is both of theoretical interest and of practical importance for the control and maintenance of networked UAV systems. In this paper, we first develop a dynamic multiplex network model for UAV networks to characterize their dynamic and heterogeneous link properties. We then propose a cross-layer fusion-based deep learning model, termed CLF-ULP, to predict future inter-UAV links based on historical topology data. 
 CLF-ULP   incorporates  graph attention networks to extract topological features within each layer and perform a cross-layer attention fusion to capture inter-layer dependencies. Furthermore, a shared-parameter long short-term memory  network is employed to model the temporal evolution of each layer. To improve embedding quality and link prediction performance, we develop a joint loss function that considers both intra-layer and inter-layer UAV  adjacency. Extensive experiments on simulated UAV datasets under diverse mobility patterns demonstrate that CLF-ULP achieves state-of-the-art performance in predicting links within dynamic multiplex UAV networks.
\end{abstract}


\section{Introduction}
\IEEEPARstart{I}{n}   recent years, Unmanned Aerial Vehicle (UAV) networks have attracted growing attention in both civilian and military domains for applications beyond the capabilities of individual UAVs, such as extending communication coverage \cite{8364586,11235946}, enabling collaborative operations \cite{9321707,9756371}, and supporting large-scale sensing \cite{10103189,10621135,11072286}. Given the diversity of tasks, UAVs must be versatile and capable of establishing various types of links with others to facilitate effective collaboration. However, these links are highly dynamic due to intrinsic UAV constraints such as limited power, wireless channels, and mobility. This dynamic and heterogeneous nature of the links adds to the complexity of UAV networks.

Link prediction in UAV networks aims to forecast future connectivity between UAVs based on current or historical information \cite{zhu2024topology}. It not only deepens our understanding of UAV network topology but also supports a range of applications, such as proactive routing, energy-efficient communication, and improved network resilience \cite{10430396}. However, this problem remains highly challenging due to the dynamic and heterogeneous nature of UAV networks.

So far, only a few link prediction methods \cite{zhang2023adaptive,10960266,long2024enhanced}, primarily adapted from mobile ad hoc networks, have been proposed specifically for UAV networks. These approaches typically rely on UAV position and mobility information to make predictions, which are then used in applications such as routing \cite{zhang2023adaptive}, clustering \cite{10960266}, and resource allocation \cite{long2024enhanced}.
However, they suffer from several limitations. First, they focus on localized link prediction tailored to specific applications, lacking a global perspective on the topology evolution of the entire UAV network. Second, they often become inapplicable when critical UAV operational information is unavailable, such as in non-cooperative or adversarial environments.

In contrast, numerous link prediction methods, ranging from heuristic graph algorithms \cite{lu2011link} to machine learning and deep learning \cite{qin2023temporal}, have been developed for general networks such as social and biological networks. However, these methods have not been systematically explored in the context of UAV networks, likely due to their inherent complexity and the lack of relevant datasets. This raises critical concerns regarding the applicability and effectiveness of these methods in UAV-specific scenarios, which are characterized by dynamic and heterogeneous link patterns.

In this paper, we investigate link prediction in complex UAV networks. The challenges of this problem are twofold. The first challenge lies in modeling complex UAV networks with dynamic and heterogeneous links. To address this, we leverage concepts from network science \cite{barabasi2013network,boccaletti2014structure} and develop a dynamic multiplex representation for complex UAV networks. Since there are no publicly available datasets for such networks, we employ typical UAV mobility models to simulate node movement and construct dynamic multiplex topologies. 

The second challenge is how to comprehensively utilize both structural and temporal information from dynamic multiplex networks to enable effective link prediction. Existing dynamic link prediction methods \cite{qin2023temporal} are not well suited to handle heterogeneous UAV links. To the best of our knowledge, there are currently no methods that can simultaneously address both dynamic and heterogeneous links in UAV networks. To tackle this, we propose a cross-layer attention fusion mechanism to jointly learn intra-layer and inter-layer topological features. We then employ a shared-parameter Long Short-Term Memory (LSTM) network \cite{hochreiter1997long} to model the temporal evolution of each layer, leveraging the inherent consistency in dynamics across layers. In addition, we introduce both intra-layer and inter-layer embedding consistency losses to guide high-quality representation learning.

Our main contributions are summarized as follows.

\begin{itemize}
\item We propose a dynamic multiplex UAV network model consisting of multiple network layers. In each layer, UAVs establish dynamic links to collaborate on specific tasks, and the layers are interconnected by links between identical UAV nodes across different layers.  Based on this model, we generate a series of benchmark datasets to facilitate research on dynamic multiplex UAV networks.
\item  We propose a deep learning-based model for link prediction in dynamic multiplex UAV networks. Our model integrates structural information within individual layers, across different layers, and temporal dynamics from topological sequences to accurately predict future links. Comprehensive experiments demonstrate that our model consistently outperforms existing methods, achieving state-of-the-art performance in link prediction. To facilitate reproducibility, the source code and datasets are publicly  available at \url{https://github.com/turtlo6/CLF-ULP}.

\end{itemize}

\begin{figure}[!t]
\centering
\includegraphics[width=1\linewidth]{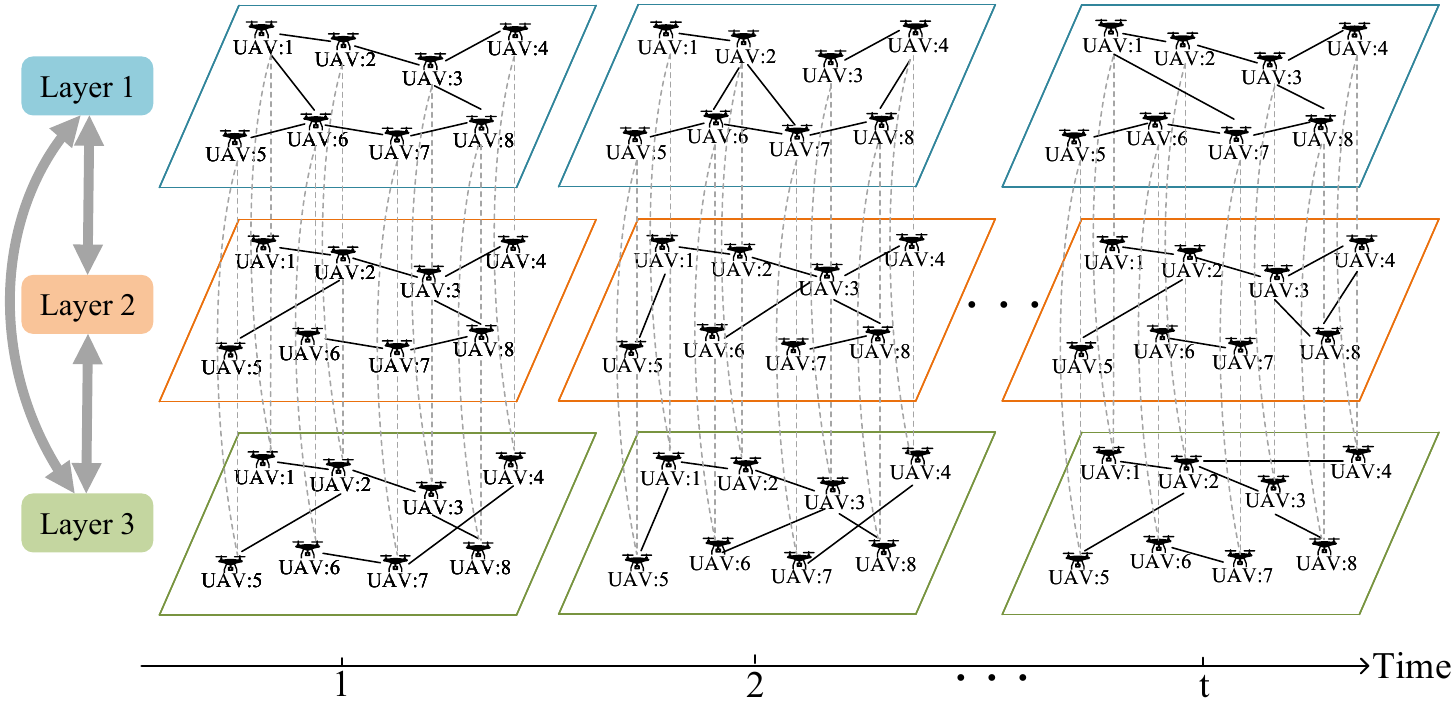}
\caption{
An example of a dynamic multiplex UAV network represented as a sequence of snapshots over time. The network consists of three layers corresponding to different link types among eight shared nodes.  Intralayer links connect different nodes within a layer, while interlayer links connect identical nodes across layers. }
\label{fig1}
\end{figure}

The remainder of the paper is organized as follows. 
Section~II reviews the related work on link prediction. 
Section~III introduces the dynamic multiplex UAV network model and formally defines the link prediction problem. 
Section~IV details the proposed link prediction model. 
Section~V presents the experimental results and analyzes the performance of the proposed prediction model.
Finally, Section~VI concludes the paper.

\section{Related work}
In this section, we provide a brief review of link prediction research in dynamic networks, multilayer networks, and  specifically in UAV networks.

\subsection{Dynamic Link Prediction}
Many real-world networks exhibit dynamic topological structures, such as UAV networks. Dynamic link prediction aims to forecast future links based on historical link information. Unlike static link prediction, it requires not only the structural information of the network but also its temporal evolution patterns to make accurate predictions.
He et al. proposed Top-Sequential-Stacking \cite{he2024sequential}, which sequentially stacks static topological features from network snapshots to construct feature vectors that incorporate both structural and temporal information for link prediction.
Sankar et al. introduced DySAT \cite{sankar2020dysat}, which utilizes a structural self-attention layer to extract features from each graph snapshot and a temporal self-attention layer to capture dynamic patterns across snapshots for learning node embeddings.
Pareja et al. proposed EvolveGCN \cite{pareja2020evolvegcn}, which adapts the Graph Convolutional Network (GCN) model along the temporal dimension and uses an Recurrent Neural Network  (RNN) to evolve the GCN parameters.
Chen et al. proposed a dynamic link prediction model called E-LSTM-D \cite{chen2019lstm}, which adopts an encoder-decoder architecture where stacked LSTM modules are integrated into the framework to effectively capture the evolving patterns of the network.

\subsection{Multilayer Link Prediction}
In multilayer link prediction, it is necessary not only to consider the network structure within each layer but also to leverage the consistency, complementarity, and coupling among different layers. Hristova et al. \cite{hristova2016multilayer} extended traditional similarity metrics such as Adamic-Adar and Jaccard coefficient from single-layer scenarios to multilayer settings. Najari et al.  \cite{najari2019link} computed Adamic-Adar and Jaccard coefficients in the target layer and combined structural information from other layers by weighted fusion of similarity scores across layers to improve prediction accuracy. Du et al. \cite{du2020cross}  proposed a Skip-gram based approach that constructs a joint embedding space allowing nodes across different networks  to share representations, thereby  enabling cross-layer information fusion. Ren et al. introduced CGNN \cite{ren2024link}, in which nodes aggregate complementary information from aligned nodes in other layers.

\subsection{Link Prediction in UAV Networks}
Early research on UAV technology primarily focused on single-UAV systems. 
With the increasing complexity of tasks, researchers began to investigate the coordination of multiple UAVs.  Hentati et al. \cite{hentati2023simulation} analyzed the evolution from single UAVs to cooperative multi-UAV systems, with particular emphasis on their architectures and applications. In addition, Chriki et al. \cite{chriki2019fanet} reviewed the routing protocols and mobility models employed in UAV networks, and summarized the key security issues that need to be addressed.
In more recent studies, UAV networks have been increasingly regarded as a type of cyber-physical system \cite{wang2019survey}, highlighting the coupling among different layers such as sensing, communication, and control. 

Existing work on link prediction in UAV networks is largely application-driven, emphasizing specific tasks such as link quality estimation \cite{cerar2021machine} and failure prediction \cite{danilchenko2023online}. These fine-grained predictions are essential for ensuring resilient routing, stable clustering, and effective resource utilization in highly dynamic and mobile UAV environments. Most approaches are adapted from the domain of mobile ad hoc networks and make use of UAV mobility and positional information. However, these methods are often localized, focusing on individual link behavior, and thus fail to capture the holistic evolution of the overall network topology. 

\section{Network model and problem formulation}
In this section, we introduce the model of a dynamic multiplex UAV network and define the link prediction task within such a network to clarify the research problem addressed in this work.
\subsection{Dynamic Multiplex UAV Networks} 
We employ modeling methods from network science \cite{barabasi2013network,boccaletti2014structure} to develop a dynamic multiplex network topology for UAV networks with dynamic and heterogeneous links. Specifically, a dynamic multiplex UAV network consists of multiple layers corresponding to different tasks such as sensing, search, and communication. Each UAV participates as a node in every layer. In each layer, the moving UAVs establish intralayer links with others within a range determined by the specific task of that layer, forming mobile ad hoc networks. The same UAV nodes are connected across layers via interlayer links. An example of a dynamic multiplex UAV network is given in Fig.~\ref{fig1}.

Let us assume a sequence of graph snapshots $\{G_1^{(l)}, G_2^{(l)}, \dots, G_T^{(l)}\}$, where $l \in [1,r]$ and $r$ denotes the number of layers. Each snapshot $G_t^{(l)} = (V^{(l)}, E_t^{(l)},X_t^{(l)})$ represents the network topology in the $l$-th layer at time step $t$, where  $E_t^{(l)}$ denotes the set of links   in the $l$-th layer at time step $t$; $V^{(l)}$ denotes the  set of nodes in the $l$-th layer, which remains fixed over time; \( X_t^{(l)} \) represents the initial node feature matrix. Since the node sets are identical across all layers and invariant over time, \( X_t^{(l)} \) remains the same for all layers at all  time steps. Without loss of generality, we define it such that each row corresponds to the \( d_x \)-bit IP address of a UAV.   The inter-layer link set is denoted by 
$E^{'} = \bigl\{ (v_i^{(l)}, v_i^{(l')}) \bigm| 
    v_i^{(l)} \in V^{(l)},\ v_i^{(l')} \in V^{(l')},\ 
    l, l' \in [1, r],\ l \ne l',\ i = 1, \dots, N \bigr\}$,
where \( v_i^{(l)} \) represents the \(i\)-th node in the \(l\)-th layer, and \(N\) is the number of nodes per layer.  
It is worth noting that all layers share the same node set, which distinguishes the proposed multiplex network from general multilayer networks, where different layers have different node sets, and nodes in different layers may be connected due to certain types of dependencies.
The adjacency matrix  of $G_t^{(l)}$ is denoted by $A_t^{(l)}$.  If there is a link between  nodes $v_i^{(l)}$ and $v_j^{(l)}$ in the $l$-th layer at time $t$, then the element at the $i$-th row and $j$-th column of $A_t^{(l)}$, denoted as $A_{t;ij}^{(l)}$, is set to 1; otherwise, $A_{t;ij}^{(l)} = 0$.

\subsection{Link prediction in Dynamic Multiplex UAV Networks}
 The task of link prediction in our work aims to simultaneously predict the future links in all layers of a dynamic multiplex UAV network. For time step \( t_0 \) and the \( l \)-th layer, given a historical snapshot sequence of length \( L \), \( \{ G_{t_0 - L + 1}^{(l)}, G_{t_0 - L + 2}^{(l)}, \dots, G_{t_0}^{(l)} \} \), the task is to predict the links in \( G_{t_0 + 1}^{(l)} \), i.e., to obtain the predicted adjacency matrix \( \hat{A}_{t_0 + 1}^{(l)} \).
 This can be formally expressed as
\begin{equation}
\label{eq1}
\hat{A}_{t_0+1}^{(l)} = \mathcal{F}\left( \{ A_{t}^{(l)} \}_{t=t_0 - L + 1}^{t_0} ; \{ X_{t}^{(l)} \}_{t=t_0 - L + 1}^{t_0} \right),
\end{equation}
where \(\mathcal{F}(\cdot)\) denotes the mapping function or model to be learned from data.
Since UAV network topology evolves over time, the input includes a sequence of historical snapshots rather than a single snapshot, as in static network link prediction \cite{lu2011link}. This allows the model to learn the temporal evolution of the network. Moreover, due to the multilayer nature of UAV networks, the input includes snapshot sequences from multiple layers, which differs from typical dynamic network link prediction \cite{qin2023temporal}. This enables the model to exploit the inter-layer coupling and consistency to enhance the performance of link prediction.
\section{Link prediction model}

In this section, we provide a comprehensive description of the proposed link prediction model for UAV networks, covering its modular architecture, optimization methodology, and computational complexity analysis.
\subsection{Model Framework}
\begin{figure*}[!t]
\includegraphics[width=1\textwidth]{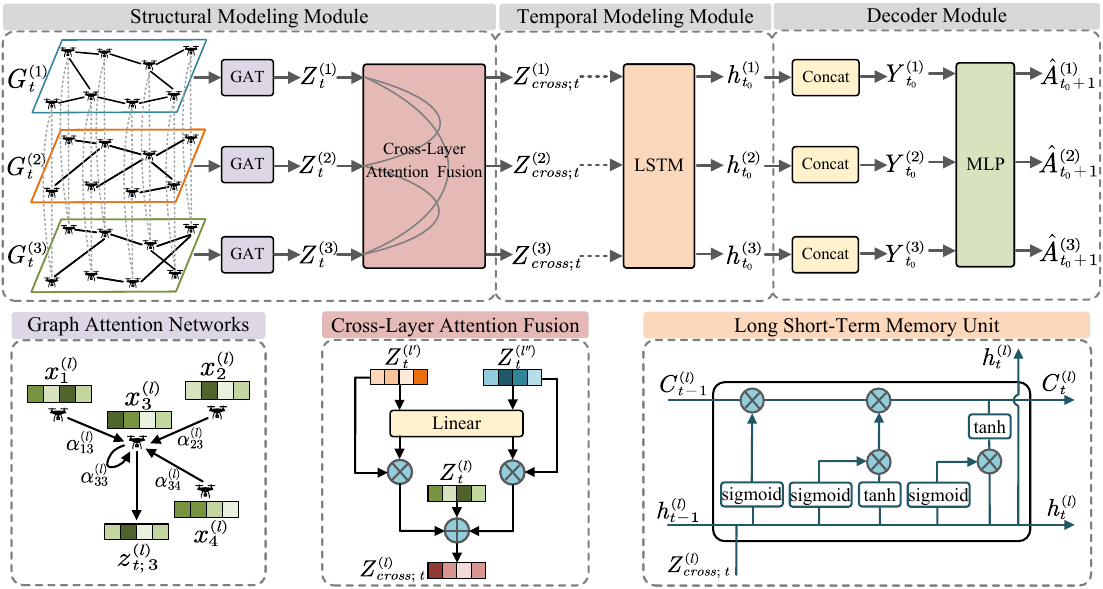}
\centering
\caption{The overall framework of the proposed link prediction model, CLF-ULP. The snapshot sequence of each layer is independently fed into a GAT to generate the corresponding intra-layer node embeddings. These embeddings are then aggregated across layers to capture inter-layer interactions through a cross-layer attention fusion mechanism. Subsequently, a shared-parameter LSTM is employed to extract temporal dependencies and produce final node embeddings that incorporate temporal information. Finally, in the decoder module,   an MLP takes as input the link features constructed by concatenating the embeddings of node pairs and transforms them into the predicted adjacency matrix.}
\label{fig3}
\end{figure*}

We propose a cross-layer fusion based UAV link prediction (CLF-ULP) model for dynamic multiplex UAV networks. CLF-ULP is a deep learning model comprising three modules: a structural modeling module, a temporal modeling module, and a decoder module, as illustrated in Fig.~\ref{fig3}. In the structural modeling module, we learn node embeddings for each layer at each historical time step. These embeddings are sequentially fed into the temporal modeling module to capture the temporal evolution patterns of node representations in each layer, generating the final temporal node features. Finally, in the decoder, the node features are concatenated to produce the final link features, which are then used to compute link scores for link prediction.
\subsubsection{Structural Modeling Module}
Graph neural networks  are often employed to capture the topological features of networks. Two popular variants of graph neural networks are GCNs \cite{zhang2019graph} and GATs \cite{velickovic2017graph}. In GCNs, the importance of neighboring nodes is determined by node degrees and remains fixed during the aggregation process. In contrast, GATs employ an attention mechanism to adaptively learn the importance of  neighbors. Given the dynamic and heterogeneous nature of UAV networks, we adopt GAT instead of GCN to learn intra-layer structural features.

 In a snapshot $G_t^{(l)}$, the attention score from node \(v_i^{(l)} \) to its neighbor node \( v_j^{(l)} \) is  defined as
\begin{equation}
\label{eq2}
e_{ij;t}^{(l)} = \sigma \left( \left(a^{(l)}\right)^T \left[ W_g^{(l)} x_{i;t}^{(l)} \| W_g^{(l)} x_{j;t}^{(l)} \right] \right),
\end{equation}
where \( x_{i;t}^{(l)} \in \mathbb{R}^{d_x} \) denotes the initial feature vector of node \( v_i^{(l)} \); \( W_g^{(l)} \in \mathbb{R}^{d_z \times d_x} \) and \( a^{(l)} \in \mathbb{R}^{2d_z} \) are learnable parameters, with $d_z$ denoting the embedding dimension; \( \| \) represents the concatenation operation and \( \sigma \) denotes a non-linear activation function, implemented as LeakyReLU in our model. The normalized attention coefficient from  node \( v_i^{(l)} \) to its neighbor node \( v_j^{(l)} \) is then calculated as 
\begin{equation}
\label{eq3}
\alpha_{ij;t}^{(l)} = \frac{\exp \left( e_{ij;t}^{(l)} \right)}{\sum\limits_{v_k^{(l)}\in  \mathcal{N}_{i;t}^{(l)}}{\exp \left( e_{ik;t}^{(l)} \right)}}, 
\end{equation}
 where \( \mathcal{N}_{i;t}^{(l)} \) denotes the set of neighbors of node \( v_i^{(l)} \) in the $l$-th layer at time $t$.
 Finally, the  embedding representation \( z_{i;t}^{(l)} \in \mathbb{R}^{d_z} \) of  node \( v_i^{(l)} \) is obtained by aggregating the transformed features of its neighbor nodes, weighted by the corresponding attention coefficients, i.e., 
\begin{equation}
\label{eq4}
z_{i;t} ^{(l)}= \sigma \left( \sum_{v_j^{(l)}\in \mathcal{N}_{i;t}^{(l)}}{\alpha_{ij;t}^{(l)}W_g^{(l)}x_{j;t}^{(l)}} \right), \ v_i^{(l)} \in V^{(l)},
\end{equation}
 where \( \sigma \) represents the ELU activation function.
By computing embeddings  for all nodes in \( G_t^{(l)} \),  we obtain  the  node embedding matrix \( Z_t^{(l)} \in \mathbb{R}^{N \times d_z} \).

However, the original GAT operates only on a single-layer graph, and thus the resulting node embeddings capture only intra-layer structural information, neglecting the inter-layer couplings present in UAV networks. To fully exploit the complementary cross-layer structural information, we propose a Cross-Layer Attention Fusion (CLAF) mechanism.

Specifically, we first employ GAT to obtain intra-layer node embeddings within each layer. Then, at each time step, for each node in layer \( l \), we aggregate information from  its corresponding nodes in the other layers via CLAF. The cross-layer attention coefficient matrix \( a^{(l' \rightarrow l)} \in \mathbb{R}^{N \times d_z} \), where each row represents the influence of a node in layer \( l' \) on its corresponding node in  layer \( l \) across different feature dimensions,  is computed as:
\begin{equation}
a^{\left( l' \rightarrow l \right)} = \sigma \left( W_{\text{cross}}^{(l)} Z^{(l')} \right), \quad l, l' \in [1, r], \quad l' \neq l,
\end{equation}
where \( Z^{(l')} \in \mathbb{R}^{N \times d_z} \) is the node embedding matrix of layer \( l' \), \( W_{\text{cross}}^{(l)} \in \mathbb{R}^{N \times N} \) is a learnable parameter matrix specific to target layer \( l \), and \( \sigma \) denotes the  sigmoid function.

Finally, the  cross-layer node embedding matrix \( Z_{\text{cross}}^{(l)} \)  is obtained by attention-weighted fusion as follows:
\begin{equation}
\label{eq6}
Z_{\text{cross}}^{(l)} = \sigma \left( Z^{(l)} + \sum_{l' \ne l} a^{(l' \rightarrow l)}\odot Z^{(l')} \right),
\end{equation}
 where $\odot$ denotes the Hadamard (element-wise) product, and \( \sigma \) denotes the ELU function.  The resulting node embeddings produced by CLAF not only capture intra-layer structural information,  but also integrate complementary information from other layers, leading to a more comprehensive representation.

\subsubsection*{2) Temporal Modeling Module}

After structural modeling, we obtain node embeddings for each layer at each time step. The temporal evolution patterns are reflected in the node embeddings across the sequence. Therefore, we employ LSTM networks for temporal modeling in UAV networks. By sequentially feeding the node embeddings at successive time steps into the LSTM units, we obtain the final feature representations that incorporate temporal information.

Each LSTM unit consists of a forget gate, an input gate, and an output gate, which collectively regulate the flow and retention of information. Given the input node embedding \( Z^{(l)}_{\text{cross};t} \in \mathbb{R}^{N \times d_z} \) of the \(l\)-th layer at time step \(t\), the operations inside the LSTM unit are defined as follows.

The forget gate determines which information from the previous hidden state should be discarded:
\begin{equation}
\label{eq7}
f_t^{(l)} = \sigma \left( W_f \left[ h_{t-1}^{(l)} \, \| \, Z^{(l)}_{\text{cross};t} \right] + b_f \right),
\end{equation}
where \( f_t^{(l)} \in \mathbb{R}^{N \times d_h} \) controls the proportion of information to forget, \( \sigma \) is the sigmoid activation function, \( W_f \) and \( b_f \) are learnable weight and bias parameters, and \( h_{t-1}^{(l)} \in \mathbb{R}^{N \times d_h} \) denotes the hidden state from the previous time step.

The input gate determines how new information is added to the cell state:

\begin{equation}
\label{eq8_1}
i_t^{(l)} = \sigma \left( W_i \left[ h_{t-1}^{(l)} \, \| \, Z^{(l)}_{\text{cross};t} \right] + b_i \right),
\end{equation}

\begin{equation}
\label{eq8_2}
\tilde{C}_t^{(l)} = \sigma \left( W_C \left[ h_{t-1}^{(l)} \, \| \, Z^{(l)}_{\text{cross};t} \right] + b_C \right).
\end{equation}
Eq. \eqref{eq8_1} computes \( i_t^{(l)} \in \mathbb{R}^{N \times d_h} \), controlling which parts of the new input should be remembered, where \( \sigma \) denotes the sigmoid activation function, and \( W_i \) and \( b_i \) are learnable weights and biases.
Eq. \eqref{eq8_2} calculates the candidate cell state \( \tilde{C}_t^{(l)} \in \mathbb{R}^{N \times d_h} \), which represents potential new information to be added to the memory, where  \( \sigma \) refers to the  \textnormal{tanh}  activation function, and \( W_C \) and \( b_C \) are learnable parameters.

By combining the outputs of the forget gate and the input gate, the updated cell state is computed as follows:

\begin{equation}
\label{eq9}
C_t^{(l)} = f_t^{(l)} \odot C_{t-1}^{(l)} + i_t^{(l)} \odot \tilde{C}_t^{(l)}.
\end{equation}

The output gate is calculated as follows:

\begin{equation}
\label{eq10_1}
o_t^{(l)} = \sigma \left( W_o \left[ h_{t-1}^{(l)} \, \| \, Z^{(l)}_{\text{cross};t} \right] + b_o \right),
\end{equation}

\begin{equation}
\label{eq10_2}
h_t^{(l)} = o_t^{(l)} \odot \sigma \left( C_t^{(l)} \right).
\end{equation}
Eq. \eqref{eq10_1} computes \( o_t^{(l)} \in \mathbb{R}^{N \times d_h} \),  controlling which information in the updated cell state should be output, where \( \sigma \) denotes the sigmoid activation function, and \( W_o \) and \( b_o \) are learnable weights and biases.
Eq. \eqref{eq10_2} computes the hidden state \( h_t^{(l)} \in \mathbb{R}^{N \times d_h} \), which serves as the output of the LSTM unit at time step \( t \), where \( \sigma \) represents the tanh activation function.

Finally, for the $l$-th layer of the UAV network, the hidden state $h_t^{(l)} \in \mathbb{R}^{N \times d_h}$ produced by the LSTM is used as the final  node representation. This representation encodes both the temporal dynamics and the structural information from both intralayer and interlayer interactions, and serves as the input to the decoder.

Given that all layers share the same set of nodes, the temporal evolution of the network topology across layers is expected to exhibit intrinsic consistency. To exploit this property, we adopt a single LSTM unit with shared parameters, into which the node embeddings from all layers are sequentially fed. This design not only reduces the number of learnable parameters but also enhances the model's ability to capture inter-layer temporal dependencies.

\subsubsection*{3) Decoder Module}

We utilize a decoder module   to predict links at time step \( t+1 \).     Specifically, for the \( l \)-th layer at time step $t$, the feature of a node pair  \( v_i^{(l)} \) and  \( v_j^{(l)} \) is obtained by concatenating their respective node embeddings:
\begin{equation}
\label{eq11}
y_{t;ij}^{(l)} = h_{t;i}^{(l)} \, \| \, h_{t;j}^{(l)},
\end{equation}
where   \( h_{t;i}^{(l)} \in \mathbb{R}^{d_h} \) is the embedding of node \( v_i^{(l)} \) at time \( t \). By applying this operation to all possible node pairs, we construct the link feature matrix \( Y_t^{(l)} \in \mathbb{R}^{N^2 \times 2d_h} \).

We employ a two-layer fully connected neural network as the decoder. The link existence probability matrix   \( S_{t+1}^{(l)} \in \mathbb{R}^{N\times N} \)  is computed as:
\begin{equation}
\label{eq12}
S_{t+1}^{(l)} = \sigma_2 \left( W_2^{(l)} \, \sigma_1 \left( W_1^{(l)} Y_t^{(l)} + b_1^{(l)} \right) + b_2^{(l)} \right),
\end{equation}
where \( W_1^{(l)} \) and \( W_2^{(l)} \) are learnable weight matrices, \( b_1^{(l)} \) and \( b_2^{(l)} \) are  bias terms, \( \sigma_1 \) denotes the \textnormal{ReLU} activation function, and \( \sigma_2 \) denotes the sigmoid activation function.

\subsection{Model Optimization}

Given sequences \(\{G_1^{(l)}, G_2^{(l)}, \dots, G_T^{(l)}\}\) for \(l \in [1, r]\), representing the snapshots of a dynamic multiplex UAV network with \(r\) layers, a sliding window of size \(L < T\) and step size 1 is applied jointly over the multiplex sequence.
Each slide extracts a multiplex subsequence of length \(L\), resulting in \(T - L + 1\) samples. In each sample, the first \(L-1\) multiplex snapshots serve as historical input, and the \(L\)-th multiplex snapshot is the prediction target.  
The collected samples are then split into training and test sets. The overall training process of the model is summarized in Algorithm~\ref{alg:1}.

Our model is trained by minimizing a loss function composed of three components: the reconstruction loss, the intra-layer consistency loss, and the inter-layer consistency loss.  
The reconstruction loss measures the discrepancy between the predicted adjacency matrix \(\hat{A}_{t_0+1}^{(l)}\) and the ground truth adjacency matrix \(A_{t_0+1}^{(l)}\).  
The structural modeling module  produces node embeddings for each layer at each time step. From a single-layer perspective, node embeddings of connected nodes should be similar, motivating the intra-layer embedding consistency loss. From a multilayer perspective, embeddings of corresponding nodes across different layers should also be similar, leading to the inter-layer embedding consistency loss.

\subsubsection*{1) Reconstruction Loss}

The \(\ell_2\) distance is commonly used to measure the discrepancy between two matrices. However, due to the sparsity of UAV networks, the number of unlinked node pairs far exceeds that of linked node pairs. Directly applying the \(\ell_2\) loss may cause the model to focus excessively on predicting non-links, thereby neglecting the importance of existing links.
To address this issue, we introduce a weighting matrix \( S^{(l)} \in \mathbb{R}^{N \times N} \), which assigns greater weights to existing links compared to non-links.  
The modified reconstruction loss \(\mathcal{L}_{\mathrm{rec}}\) is computed as:
\begin{equation}
\label{eq13_1}
\mathcal{L}_{\mathrm{rec}} = 
\sum_{l=1}^r \sum_{i=1}^N \sum_{j=1}^N 
S_{ij}^{(l)} \left( \hat{A}_{t;ij}^{(l)} - A_{t;ij}^{(l)} \right)^2,
\end{equation}
where
\begin{equation}
\label{eq13_2}
S_{ij}^{(l)} = 
\begin{cases}
\varepsilon, & A_{t;ij}^{(l)} = 1, \\
1, & A_{t;ij}^{(l)} = 0,
\end{cases}
\end{equation}
with \( S_{ij}^{(l)} \) denoting the \((i,j)\)-th entry of \( S^{(l)} \), and \(\varepsilon > 1\) a hyperparameter that amplifies the penalty on prediction errors involving existing links.
A smaller reconstruction loss indicates that the predicted adjacency matrix is closer to the ground truth.

\subsubsection*{2) Intra-layer Embedding Consistency Loss}

The embeddings of two connected nodes within a layer should be close. To achieve this, we minimize the \( \ell_2 \) norm of their difference. The intra-layer embedding consistency loss is defined as follows:
\begin{equation}
\label{eq14}
\mathcal{L}_{\mathrm{intra}} = 
\sum_{t=1}^L \sum_{l=1}^r \sum_{(v_i^{(l)}, v_j^{(l)}) \in E_t^{(l)}}
\left\| z_{t;i}^{(l)} - z_{t;j}^{(l)} \right\|_2^2,
\end{equation}
where \( z_{t;i}^{(l)} \in \mathbb{R}^{d_z} \) denotes the embedding of node \( v_i^{(l)} \) in the \( l \)-th layer at time \( t \). The summation is taken over all connected node pairs  in the \( l \)-th layer at time \( t \), and \( L \) denotes the length of the historical window  considered for link prediction. A smaller intra-layer loss indicates that embeddings of connected node pairs within a layer are closer.

\begin{algorithm}[!t]
\caption{Training Procedure of CLF-ULP}
\label{alg:1}
\begin{algorithmic}[1]
\REQUIRE Training snapshot sequence \(\{G_1^{(l)}, G_2^{(l)}, \dots, G_{N_0}^{(l)}\}\), number of training snapshots \(N_0\), number of training epochs \(K\), window size \(L\).
\ENSURE Trained model parameter \(\Theta_*\).
\STATE Initialize model parameter \(\Theta\) randomly.
\FOR{\(epoch = 1\) to \(K\)}
    \FOR{\(n = L\) to \(N_0\)}
        \STATE Construct input sequence \(\{G_{n-L+1}^{(l)}, \ldots, G_{n-1}^{(l)}\}\).
        \FOR{\(i = n-L+1\) to \(n-1\)}
            \STATE Feed \(G_i^{(l)}\) into the structural modeling module.
            \STATE Compute \(Z_{cross;i}^{(l)}\) using Eqs.~\eqref{eq2}--\eqref{eq6}.
        \ENDFOR
        \STATE Feed \(\{Z_{cross;n-L+1}^{(l)}, \ldots, Z_{cross;n-1}^{(l)}\}\) into the temporal modeling module.
        \STATE Compute \(h_{n-1}^{(l)}\) using Eqs.~\eqref{eq7}--\eqref{eq10_2}.
        \STATE Feed \(h_{n-1}^{(l)}\) into the decoder module.
        \STATE Compute \(S_n^{(l)}\) using Eqs.~\eqref{eq11}--\eqref{eq12}.
        \STATE Backpropagate the loss and update parameters \(\Theta\).
    \ENDFOR
\ENDFOR
\STATE Return optimized parameter \(\Theta_*\).
\end{algorithmic}
\end{algorithm}

\subsubsection*{3) Inter-layer Embedding Consistency Loss}

Since each node is connected to its counterparts in other layers, their representations are expected to be similar. To encourage consistency across layers, we use the \( \ell_2 \) norm to minimize the difference between the embeddings of identical nodes across layers. The inter-layer embedding consistency loss is defined as follows:
\begin{equation}
\label{eq15}
\mathcal{L}_{\text{inter}} =
\sum_{t=1}^L \sum_{1 \leq l_1 < l_2 \leq r} \sum_{i=1}^N
\left\| z_{t;i}^{(l_1)} - z_{t;i}^{(l_2)} \right\|_2^2,
\end{equation}
where \( z_{t;i}^{(l)} \) denotes the embedding of node \( i \) at time \( t \) in layer \( l \), and \( l_1 \), \( l_2 \) represent any two distinct layers. The summation is taken over all identical nodes across all layer pairs.  
A smaller inter-layer loss indicates that the embeddings of corresponding nodes across different layers are more similar.

 In summary, the overall loss function of the model is defined as:
\begin{equation}
\label{eq16}
\mathcal{L} = \mathcal{L}_{\mathrm{rec}} + \alpha \mathcal{L}_{\mathrm{intra}} + \beta \mathcal{L}_{\mathrm{inter}},
\end{equation}
where \(\alpha\) and \(\beta\) are hyperparameters that control the relative contributions of the intra-layer and inter-layer consistency losses, respectively.


\begin{table*}[!t]
\renewcommand{\arraystretch}{1.2}
\caption{UAV Network Simulation Parameter Settings}
\label{tab1}
\centering
\begin{tabularx}{\textwidth}{>{\centering\arraybackslash}X>{\centering\arraybackslash}X>{\centering\arraybackslash}X>{\centering\arraybackslash}X>{\centering\arraybackslash}X>{\centering\arraybackslash}X>{\centering\arraybackslash}X}
\hline
\makecell{Number of\\ UAVs} & \makecell{Simulation Area\\ Size ($\text{km}^2$)} & \makecell{Flight\\ Speed (m/s)} & \makecell{Communication\\ Radius (m)} & \makecell{Simulation\\ Duration (s)} & \makecell{Sampling\\ Interval (s)} \\
\hline
100 & 100 & 25$\sim$35 & 1000$\sim$3000 & 160 & 2 \\
\hline
\end{tabularx}
\vspace{-6pt}
\end{table*}

\begin{table}[!t]
\renewcommand{\arraystretch}{1.2}
\setlength{\tabcolsep}{3pt}
\caption{Statistics of UAV Network Snapshot Sequences}
\label{tab2}
\centering
\begin{tabularx}{0.48\textwidth}{c|>{\centering\arraybackslash}X>{\centering\arraybackslash}X>{\centering\arraybackslash}X>{\centering\arraybackslash}X}
\hline
Dataset & Min. Edges & Max. Edges & Avg. Edges & Avg. Density \\
\hline
RW & 430 & 522 & 476 & 0.0962 \\
GM & 683 & 789 & 739 & 0.1494 \\
RPG & 577 & 878 & 709 & 0.1478 \\
MG & 405 & 468 & 431 & 0.0873 \\
\hline
\end{tabularx}
\setlength{\tabcolsep}{6pt}
\end{table}

\subsection{Model Complexity Analysis}

The complexity of our model,  CLF-ULP,  arises from three main components: the structural modeling module, the temporal modeling module, and the decoder module.

For the structural modeling, the complexity can be analyzed in two parts. The first part is the complexity of applying GAT to each  layer at each  time step. Let \(m\) denote the average number of links in a layer at a time step. The complexity of GAT for a layer  is \(O\big(L(N d_x d_z + m d_z)\big)\),  where $L$ is the number of historical snapshots,  \(N\) is the number of nodes, \(d_x\) is the input feature dimension, and \(d_z\) is the node embedding dimension. 

The second part is the complexity of the cross-layer attention fusion. Since attention is computed via linear projections, the complexity  for each layer is \(O(L N d_z)\).

For the temporal modeling,  the complexity mainly comes from the LSTM, which is \(O\big(L d_h (d_h + d_z)\big)\), where \(d_h\) is the hidden state dimension.

Finally,  in the decoder module, a two-layer fully connected neural network  evaluates all $N^2$ node pairs, with a  corresponding  complexity  of \(O(N^2 d_h)\).

\section{Performance evaluation}

In this section, we conduct extensive experiments to evaluate the performance of the proposed CLF-ULP model. We begin by introducing the datasets, baseline methods, and evaluation metrics. Then, we compare the performance of CLF-ULP with various baselines and analyze the effects of sampling interval and UAV flight speed. Additional results, including the visualization of node embeddings, are available at \url{https://github.com/turtlo6/CLF-ULP}.

\subsection{Datasets}
We use four commonly adopted UAV mobility models to generate diverse UAV mobility datasets. A link is established between two UAVs if their distance is within a specified range. Without loss of generality, we construct three network layers in our experiments to simulate different collaboration scenarios. A mobility model is used to independently generate each layer with a distinct connection range, corresponding to different task requirements. Identical nodes across different layers are then interconnected to establish inter-layer links. The simulation parameters are summarized in Table \ref{tab1}, and the statistics of the generated network snapshot sequences are shown in Table \ref{tab2}.
The four mobility models are briefly described as follows:

Random Walk (RW) model \cite{sharma2019random} is commonly used to simulate the unpredictable movements of natural entities. In the context of UAV networks, it models UAVs as entities that randomly select directions and speeds at each time interval and maintain their selected motion for a specified duration.

Gauss-Markov (GM) model \cite{biomo2014enhanced} updates UAV trajectories based on historical speed and direction, producing smooth and continuous movement suitable for stable coverage tasks.

Reference Point Group (RPG) model \cite{hong1999group} simulates UAVs moving around a common reference point, which periodically shifts to new target locations. This model is suitable for cooperative scenarios, such as agricultural monitoring and management.

Manhattan Grid (MG) model \cite{gapeyenko2021line} simulates UAVs moving along a city-like grid layout, where movement is constrained to horizontal and vertical paths. At intersections, UAVs make probabilistic decisions to either turn or continue moving straight. This model is well-suited for simulating urban patrol and traffic monitoring scenarios.

\subsection{Baseline Methods}

To assess the performance of CLF-ULP, we compare it with several popular and recently proposed baseline methods:

Node2Vec \cite{grover2016node2vec} is a classical static graph embedding method that employs biased random walks and the Skip-gram model to learn node embeddings. It achieves a balance between local and global structural features.

EvolveGCN \cite{pareja2020evolvegcn} is a dynamic graph learning model, where a GCN is employed as a feature extractor and its parameters are dynamically updated by an RNN. Two variants of the model are proposed: EvolveGCN-O and EvolveGCN-H.

E-LSTM-D \cite{chen2019lstm}  is an end-to-end deep learning model for dynamic link prediction. It uses stacked LSTM units within an encoder-decoder framework to reconstruct evolving graph structures.

Top-Sequential-Stacking \cite{he2024sequential}  is a topological feature-based framework that sequentially stacks static topological features across graph snapshots for dynamic link prediction.

These baseline methods have not previously been applied to dynamic multiplex networks. Therefore, we adapt them to enable their application in dynamic multiplex UAV network settings. For static methods such as Node2Vec, we predict the next snapshot $G_{t_0+1}$ based on the current graph $G_{t_0}$. For dynamic models, we perform link prediction independently on each network layer and report the average performance across all layers.

\subsection{Evaluation Metrics}

We adopt two widely used evaluation metrics: the Area Under the ROC Curve (AUC) and the Average Precision (AP), where higher values indicate better performance.
AUC evaluates the model’s ability to distinguish between positive and negative classes across all threshold settings. It offers a threshold-independent assessment of model performance and is particularly robust to class imbalance.
AP, on the other hand, accounts for both precision and recall across varying thresholds. In link prediction tasks, AP places greater emphasis on the accurate identification of existing links compared to AUC, which is especially important in sparse UAV networks. Therefore, using both AUC and AP provides a more comprehensive evaluation of model performance.

\begin{table*}[!t]
\renewcommand{\arraystretch}{1.2}
\caption{The results of AUC and AP for All the Methods Across the Four Mobility Models}
\label{tab3}
\centering
\begin{tabularx}{\textwidth}{c|>{\centering\arraybackslash}X>{\centering\arraybackslash}X>{\centering\arraybackslash}X>{\centering\arraybackslash}X>{\centering\arraybackslash}X>{\centering\arraybackslash}X>{\centering\arraybackslash}X>{\centering\arraybackslash}X}
\hline
\multirow{2}{*}{Methods} & \multicolumn{2}{c}{RW} & \multicolumn{2}{c}{GM} & \multicolumn{2}{c}{RPG} & \multicolumn{2}{c}{MG}  \\ 
\cline{2-9}
                         & AUC    & AP            & AUC    & AP            & AUC    & AP             & AUC    & AP             \\ 
\hline
Node2Vec                 & 0.5880 & 0.1077        & 0.5931 & 0.1659        & 0.5338 & 0.1581         & 0.5597 & 0.0921         \\
EvolveGCN-O              & 0.8327 & 0.3401        & 0.7967 & 0.3998        & 0.7888 & 0.3608         & 0.8459 & 0.3232         \\
EvolveGCN-H              & 0.8518 & 0.3708        & 0.8193 & 0.4379        & 0.7716 & 0.3452         & 0.8513 & 0.3213         \\
Top-Sequential-Stacking  & 0.9325 & 0.5098        & 0.8647 & 0.3862        & 0.6724 & 0.2295         & 0.9050 & 0.3439         \\
E-LSTM-D                 & 0.9547 & 0.6563        & 0.9440 & 0.6857        & 0.8489 & 0.5147         & 0.9594 & 0.6452         \\
CLF-ULP                  & \textbf{0.9790} & \textbf{0.6650}        & \textbf{0.9748} & \textbf{0.6951}        & \textbf{0.9470} & \textbf{0.5853}         & \textbf{0.9820} & \textbf{0.6591}         \\
\hline
\end{tabularx}
\end{table*}

\subsection{Parameter Settings}

With the simulation parameters provided in Table~\ref{tab1}, we use BonnMotion \cite{aschenbruck2010bonnmotion}, a mobility scenario generation and analysis tool, to generate UAV mobility datasets under different mobility models.
For each mobility model, 80 network snapshots are collected at 2-second intervals over a 160-second period.
Using a sliding window of size 11 and a stride of 1, we generate 70 samples, with the first 50 used for training and the remaining 20 for testing. In each sample, the first 10 snapshots are used as input to predict the 11th snapshot.
All methods employ 128-dimensional node embeddings, except for Top-Sequential-Stacking, whose embedding dimension depends on the number of employed topological features. For CLF-ULP, we use a two-layer GAT for structural modeling, a single-layer LSTM for temporal modeling, and a two-layer MLP for decoding. Additional hyperparameters are provided along with our released code  at \url{https://github.com/turtlo6/CLF-ULP}. The baseline models adopt the parameter settings specified in their original papers and official implementations.

\begin{figure*}[!t]
\includegraphics[width=1\textwidth]{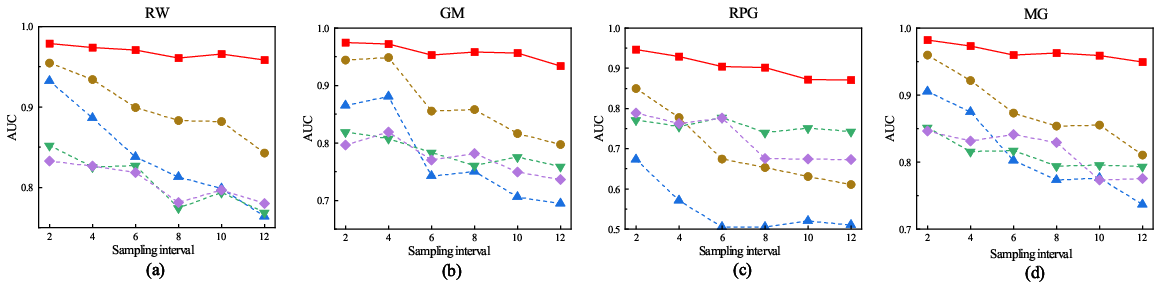}
\vspace{-12pt}
\end{figure*}
\begin{figure*}[!t]
\vspace{-12pt}
\includegraphics[width=1\textwidth]{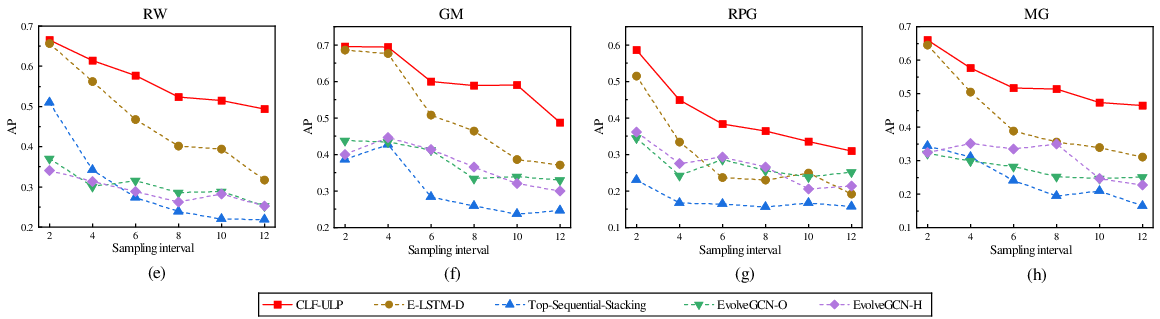}
\caption{AUC and AP vs. sampling interval for different methods across the four mobility models.}
\vspace{-10pt}
\label{fig4}
\end{figure*}

\subsection{Experimental Results}

\subsubsection*{1) Comparison with Baselines}
Table~\ref{tab3} presents the performance of various link prediction methods under different UAV mobility models. The proposed CLF-ULP model achieves the highest AUC and AP scores across all mobility scenarios.

As a static graph embedding method, Node2Vec lacks the ability to capture the evolving dynamics of UAV networks, leading to consistently poor performance across all mobility models. EvolveGCN-O and EvolveGCN-H fail to account for the inherent sparsity of UAV networks and tend to overemphasize non-existent links, which ultimately results in significantly lower AP scores.
E-LSTM-D employs MLPs rather than GNNs to extract structural information, which limits its ability to effectively capture node connectivity. While it performs relatively well under simpler mobility models such as RW, GM, and MG, its performance degrades in more complex scenarios like RPG.
Top-Sequential-Stacking relies on handcrafted feature combinations, making its performance highly sensitive to feature selection and resulting in instability across different datasets. Under the RPG scenario, both its AUC and AP scores remain relatively low.

In contrast, the proposed CLF-ULP model is specifically designed for dynamic multiplex UAV networks. It effectively learns structural features and captures temporal dependencies. In particular, it integrates node embeddings across layers through a cross-layer attention fusion mechanism. These capabilities enable CLF-ULP to achieve superior performance in modeling and predicting the structures of dynamic multiplex UAV networks.

\subsubsection*{2) Impact of Sampling Interval}
In Table~\ref{tab1}, the sampling interval is set to 2 seconds. However, the length of the sampling interval may influence link prediction performance. Longer intervals between consecutive graph snapshots weaken temporal dependencies, making it more difficult to capture network evolution patterns.
To investigate the impact of the sampling interval, we generate additional datasets by varying the sampling interval and the total simulation duration, while keeping all other parameters fixed. Fig. \ref{fig4} shows the performance of five methods under different sampling intervals, excluding Node2Vec due to its significantly inferior performance.
As the sampling interval increases, prediction performance generally declines across all mobility models, with the most significant drop observed under the RPG scenario. Nevertheless, the proposed CLF-ULP model exhibits a slower performance degradation compared to other methods and consistently achieves the best results across all sampling intervals. This demonstrates that CLF-ULP is more robust to weakened temporal dependencies between graph snapshots, highlighting its superior adaptability to varying sampling intervals.

\begin{figure*}[!t]
\vspace{-10pt}
\includegraphics[width=1\textwidth]{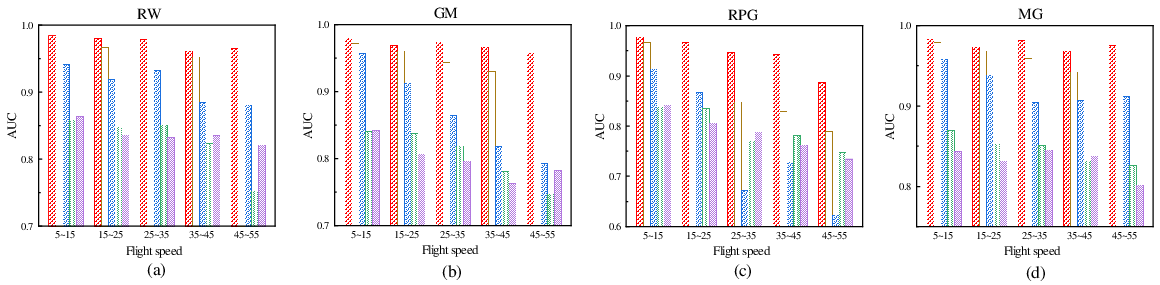}
\vspace{-12pt}
\end{figure*}
\begin{figure*}[!t]
\vspace{-12pt}
\includegraphics[width=1\textwidth]{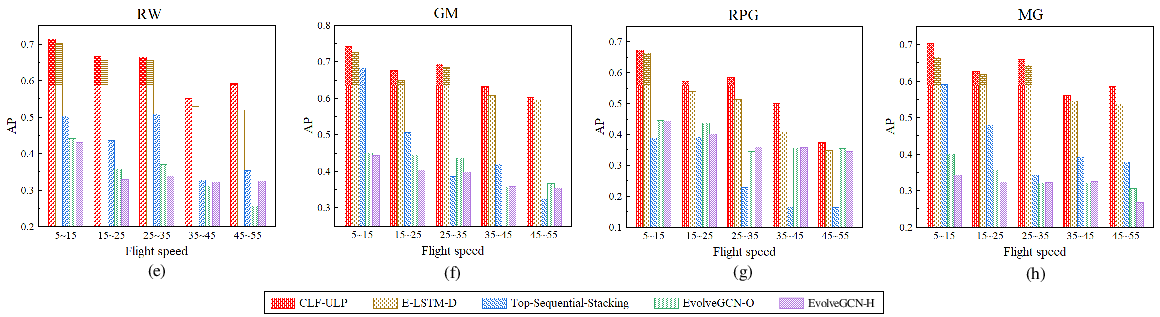}
\caption{AUC and AP vs. flight speed for different methods across the four mobility models.}
\label{fig5}
\end{figure*}


\subsubsection*{3) Impact of Flight Speed}

In Table~\ref{tab1}, the UAV flight speed is fixed within the range of 25 to 35 m/s. However, UAV speeds may vary significantly depending on the specific application scenario. At lower speeds, the network topology evolves more slowly, while at higher speeds, it changes more rapidly.
To evaluate the impact of flight speed, we generate additional datasets by varying the UAV speed range while keeping all other parameters fixed. Fig.~\ref{fig5} presents a performance comparison of different methods under various speed settings.   Node2Vec is again excluded due to its significantly  inferior performance.
 As the flight speed increases, the prediction performance of all methods generally exhibits a downward trend with noticeable fluctuations. Among them,  CLF-ULP demonstrates  minimal performance variation, highlighting its robustness, while Top-Sequential-Stacking experiences the most pronounced fluctuations. Overall, CLF-ULP consistently outperforms the baselines across all speed settings.
\subsubsection*{4) Ablation Study on Inter-layer Modeling}
To verify the effectiveness of incorporating inter-layer relationships in link prediction, we design a comparison method, referred to as No Inter-Layer, which retains the overall architecture of CLF-ULP but removes all components related to inter-layer interactions.
Specifically, in the structural modeling, GAT is applied to each layer independently, without inter-layer attention fusion. In the temporal modeling, each layer employs an independent LSTM instead of a shared-parameter LSTM. During training, the inter-layer consistency loss is also removed from the loss function.

The comparative results  are presented in Table~\ref{tab4}. The results indicate that the No Inter-Layer variant experiences a noticeable performance drop, particularly in terms of the AP metric. This supports our main claim that incorporating inter-layer relationships is crucial for effective link prediction in dynamic multiplex UAV networks.
\begin{table*}[htbp]
\renewcommand{\arraystretch}{1.2}
\caption{AUC and AP with and without considering inter-layer relationships}
\label{tab4}
\centering
\begin{tabularx}{\textwidth}{c|>{\centering\arraybackslash}X>{\centering\arraybackslash}X>{\centering\arraybackslash}X>{\centering\arraybackslash}X>{\centering\arraybackslash}X>{\centering\arraybackslash}X>{\centering\arraybackslash}X>{\centering\arraybackslash}X}
\hline
\multirow{2}{*}{Methods} & \multicolumn{2}{c}{RW} & \multicolumn{2}{c}{GM} & \multicolumn{2}{c}{RPG} & \multicolumn{2}{c}{MG}  \\ 
\cline{2-9}
                         & AUC    & AP            & AUC    & AP            & AUC    & AP             & AUC    & AP             \\ 
\hline
No Inter-Layer  & 0.9723 & 0.6106        & 0.9698 & 0.6522        & 0.9293 & 0.5223         & 0.9772 & 0.6150         \\
CLF-ULP                  & {0.9790} & {0.6650}        & {0.9748} & {0.6951}        & {0.9470} & {0.5853}         & {0.9820} & {0.6591}         \\
\hline
\end{tabularx}
\end{table*}
 
\subsubsection*{5) Node Embedding Visualization}
 A key novelty of our model is its incorporation of inter-layer relationships by aggregating the node embeddings across layers through attention fusion to obtain enhanced node representations.

\begin{figure}[!t]
\centering
\includegraphics[width=1\linewidth]{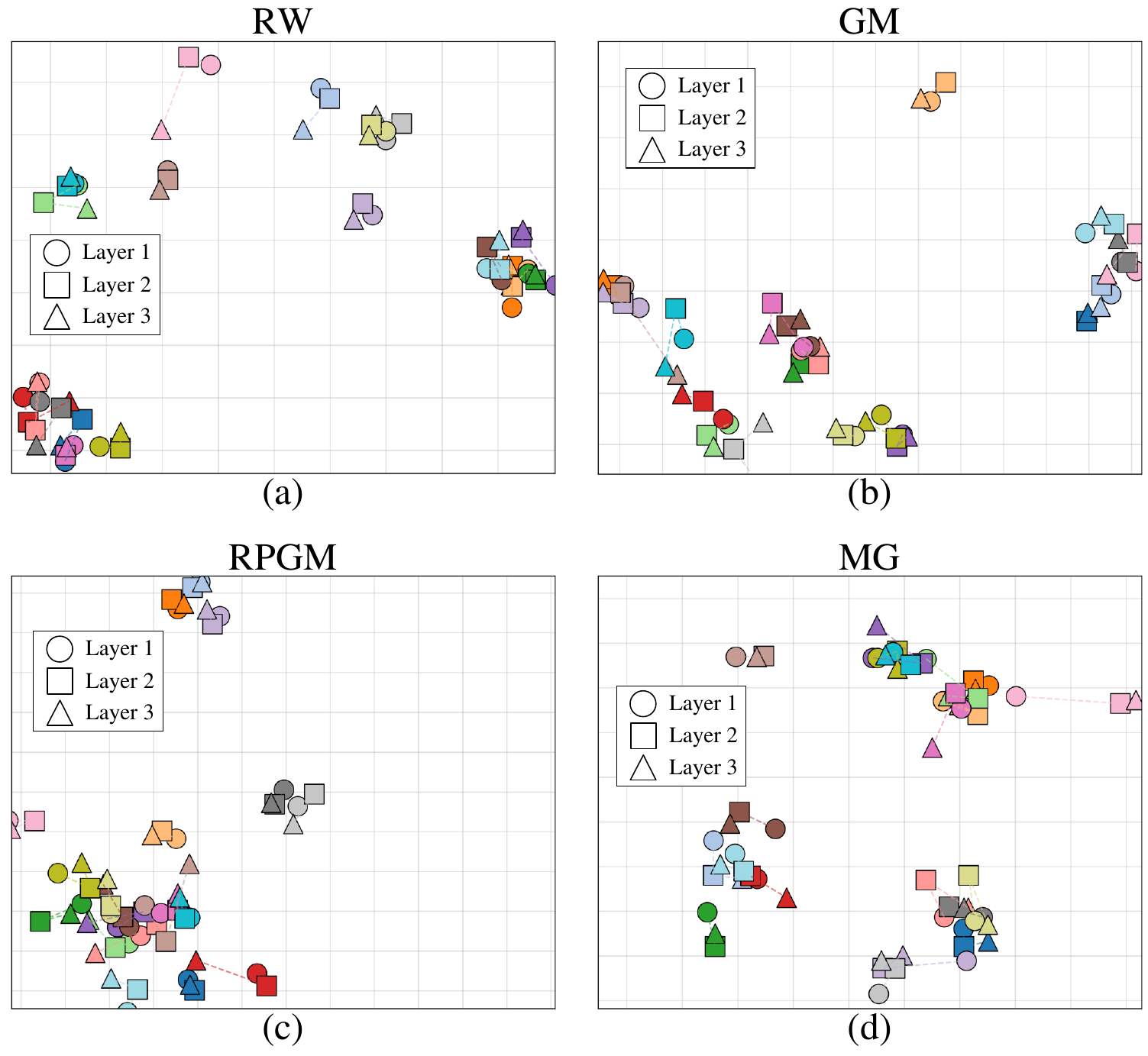}
\caption{Node embedding visualization for the four mobility models.}
\label{fig_6}
\end{figure}

We randomly select a multiplex snapshot from the dataset for each mobility model and apply t-SNE \cite{maaten2008visualizing} to project the node embeddings into a two-dimensional space. In this space, the closer two points are, the more similar their embeddings. As shown in Fig.~\ref{fig_6}, we visualize the embeddings of the first 20 nodes, sorted by label, and omit the rest for clarity.

We observe that markers with different shapes but the same color tend to cluster together, indicating that identical nodes across different layers have similar embeddings. This suggests that the node representations generated by CLF-ULP effectively capture inter-layer relationships, thereby contributing to improved link prediction performance.

\section{Conclusion}

In summary, we address the problem of link prediction in dynamic multiplex UAV networks. Specifically, we propose a dynamic multiplex network architecture comprising multiple layers, each representing a distinct type of interaction among UAV nodes. Furthermore, we introduce a deep learning-based model, CLF-ULP, for link prediction in such networks. The model captures both intra-layer structural features and inter-layer dependencies through a cross-layer attention fusion of GAT-derived embeddings,  while temporal dynamics are modeled using a shared-parameter LSTM.
Extensive experiments under various UAV mobility models show that CLF-ULP consistently outperforms baseline methods and exhibits greater robustness to variations in sampling intervals and flight speeds. 
Ablation studies and node embedding visualizations further provide explanatory insights into the superiority of CLF-ULP.
We believe that the proposed link prediction framework holds potential for broader application to other types of dynamic multiplex networks beyond UAV scenarios.

\bibliographystyle{IEEEtran}
\bibliography{IEEEabrv,IEEEexample}

\begin{IEEEbiography}[{\includegraphics[width=1in,height=1.25in,clip,keepaspectratio]{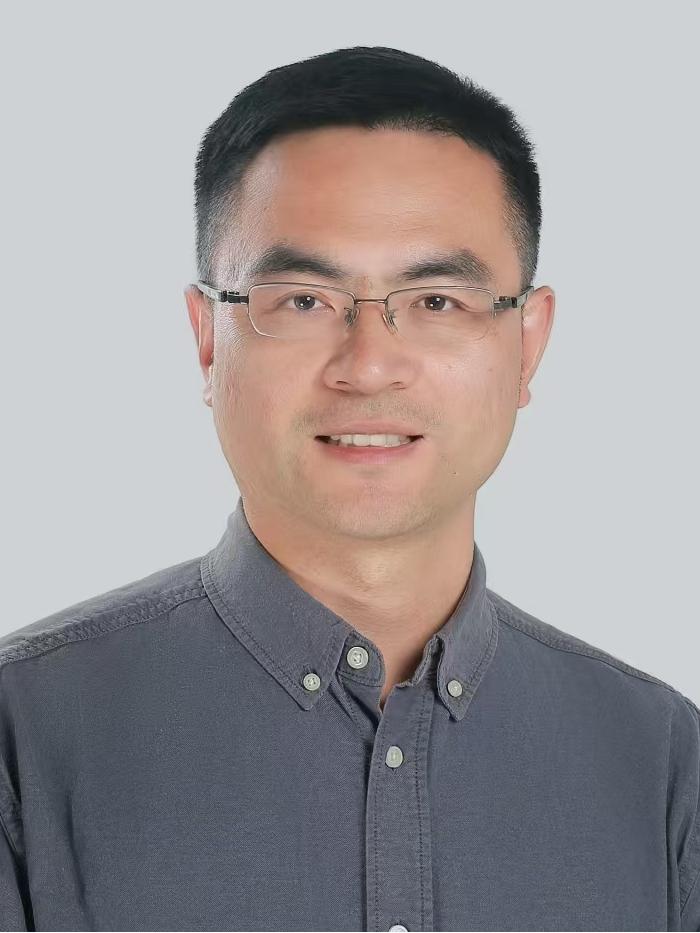}}]{Cunlai Pu}
received the Ph.D. degree in information and communication engineering from Southeast University, Nanjing, China, in 2012. He is currently an Associate Professor with the School of Computer Science and Engineering, Nanjing University of Science and Technology, Nanjing, China. His interests include network science, communication systems, and network optimization.
\end{IEEEbiography}

\vspace{-20pt}
\begin{IEEEbiography}[{\includegraphics[width=1in,height=1.25in,clip,keepaspectratio]{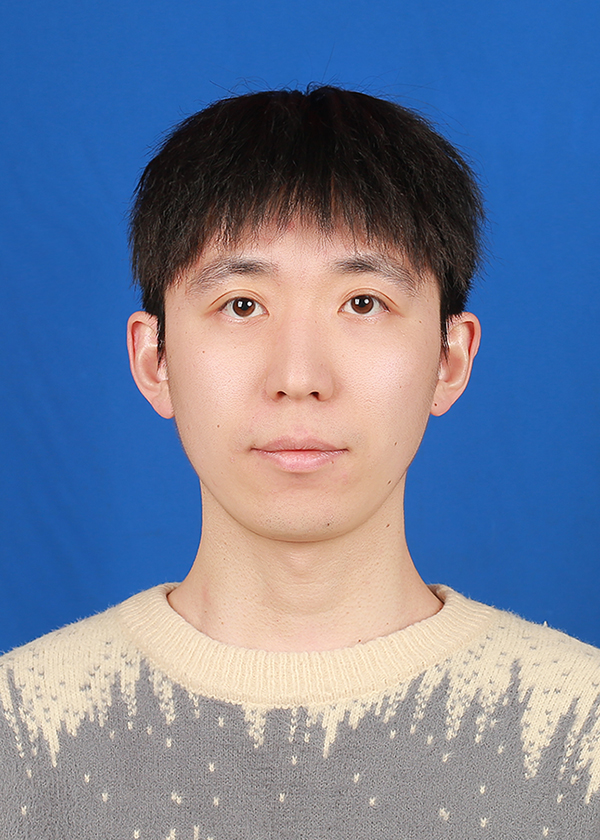}}]{Fangrui Wu}
received the B.S. degree from Guangdong University of Technology, Guangzhou, China, in 2023. He is currently working toward the M.S. degree in computer science with the School of Computer Science and Engineering, Nanjing University of Science and Technology. His research interests include dynamic link prediction and graph machine learning.
\end{IEEEbiography}

\vspace{-20pt}
\begin{IEEEbiography}[{\includegraphics[width=1in,height=1.25in,clip,keepaspectratio]{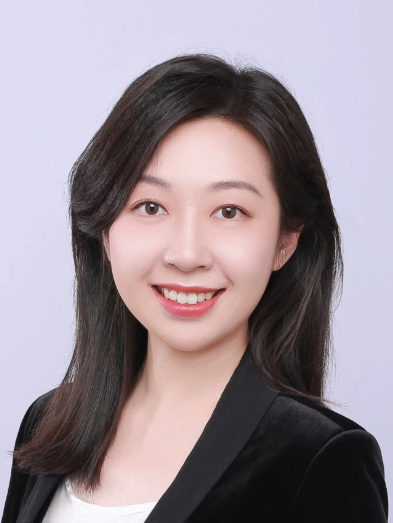}}]{Zhe Wang}
 received the Ph.D. degree in electrical engineering from The University of New South Wales, Sydney, NSW, Australia, in 2014. From 2014 to 2020, she was a Research Fellow with The University of Melbourne, Melbourne, VIC, Australia, and Singapore University of Technology and Design, Singapore. She is currently a Professor with the School of Computer Science and Engineering, Nanjing University of Science and Technology, Nanjing, China. Her research interests include applications of optimization, reinforcement learning, and game theory in communications and networking. She is the Editor of IEEE Open Journal of the Communications Society.
\end{IEEEbiography}

\vspace{-20pt}
\begin{IEEEbiography}[{\includegraphics[width=1in,height=1.25in,clip,keepaspectratio]{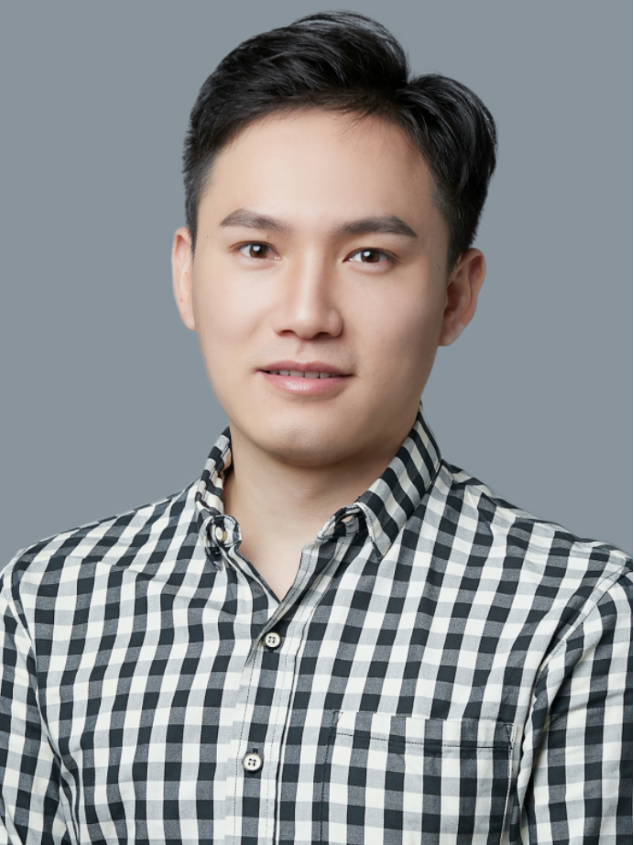}}]{Xiangbo Shu}
(Senior Member, IEEE) received the Ph.D. degree from  Nanjing University of Science and Technology in 2016. From 2014 to 2015, he was a Visiting Scholar with  National University of Singapore, Singapore. He is currently a Professor with the School of Computer Science and Engineering, Nanjing University of Science and Technology, China. His current research interests include computer vision, multimedia and machine learning. He has authored over 100 journals and conference papers in these areas, including IEEE TPAMI, TNNLS, TIP; CVPR, ICCV, ECCV, and ACM MM, etc. He has served as the editorial boards of the IEEE TNNLS, and the IEEE TCSVT. He is also the Member of ACM, the Senior Member of CCF, and the Senior Member of IEEE.
\end{IEEEbiography}

\end{document}